\documentclass[runningheads]{llncs}
\usepackage{graphicx}
\usepackage{booktabs}
\usepackage{multirow}
\usepackage{tabularx,ragged2e}
\newcolumntype{C}{>{\Centering\arraybackslash}X}
\usepackage{amssymb}
\usepackage{amsmath}

\DeclareMathOperator*{\argmin}{arg\,min}
\usepackage{pifont}
\usepackage{microtype}
\usepackage{cite}
\usepackage[colorlinks=true]{hyperref}

\begin{document}
\title{Information-based Disentangled Representation Learning for Unsupervised MR Harmonization}
\titlerunning{Disentangled Representation Learning for Unsupervised MR
Harmonization}


\author{
Lianrui~Zuo\inst{1,2}\orcidID{0000-0002-5923-9097} \and
Blake~E.~Dewey\inst{1}\orcidID{0000-0003-4554-5058} \and
Aaron~Carass\inst{1}\orcidID{0000-0003-4939-5085} \and
Yihao~Liu\inst{1}\orcidID{0000-0003-3187-9903} \and
Yufan~He\inst{1}\orcidID{0000-0003-4095-9104} \and
Peter~A.~Calabresi\inst{3}\orcidID{0000-0002-7776-6472} \and
Jerry~L.~Prince\inst{1}\orcidID{0000-0002-6553-0876}
}
\authorrunning{L.~Zuo~et~al.}

\institute{
Department of Electrical and Computer Engineering, \\
Johns Hopkins University, Baltimore, MD~21218,~USA \\
\email{\{lr\_zuo, blake.dewey, aaron\_carass, yliu236, heyufan, prince\}@jhu.edu} \and 
Laboratory of Behavioral Neuroscience, National Institute on Aging, \\
National Institute of Health, Baltimore, MD~20892,~USA \and
Department of Neurology, \\
Johns Hopkins School of Medicine, Baltimore, MD~21287,~USA \\
\email{pcalabr1@jhmi.edu}
}

%
%
\maketitle          
\begin{abstract}
Accuracy and consistency are two key factors in computer-assisted magnetic resonance~(MR) image analysis. 
However, contrast variation from site to site caused by lack of standardization in MR acquisition impedes consistent measurements. 
In recent years, image harmonization approaches have been proposed to compensate for contrast variation in MR images. 
Current harmonization approaches either require cross-site traveling subjects for supervised training or heavily rely on site-specific harmonization models to encourage harmonization accuracy. 
These requirements potentially limit the application of current harmonization methods in large-scale multi-site studies. 
In this work, we propose an unsupervised MR harmonization framework, CALAMITI~(Contrast Anatomy Learning and Analysis for MR Intensity Translation and Integration), based on information bottleneck theory. 
CALAMITI learns a disentangled latent space using a unified structure for multi-site harmonization without the need for traveling subjects. 
Our model is also able to adapt itself to harmonize MR images from a new site with fine tuning solely on images from the new site. 
Both qualitative and quantitative results show that the proposed method achieves superior performance compared with other unsupervised harmonization approaches. 

 \end{abstract}

\section{Introduction}
Magnetic resonance~(MR) imaging is a commonly used non-invasive imaging modality due to its flexibility and good tissue contrast. 
For the purposes of describing MR imaging analytically, we can think of an MR image as a function~(i.e., imaging equation) of the anatomy being imaged and the associated acquisition parameters~\cite{brown2014magnetic}. 
By changing the acquisition parameters or underlying imaging equations, MR images with different contrasts can be generated. 
To take advantage of this flexibility, MR images of the same anatomy with different contrasts are often acquired in a single session. 
For example, T$_1$-weighted~(T$_1$-w) images are typically used to achieve balanced contrast between T$_2$-weighted~(T$_2$-w) images~\cite{brown2014magnetic}. 
However, a consequence of this flexibility is that there is no standardization when it comes to MR contrasts. 
For example, both magnetization-prepared rapid gradient echo~(MPRAGE) and spoiled gradient echo~(SPGR) are commonly used T$_1$-w images with very different visual appearances. 
This lack of standardization makes machine learning~(ML) models trained on MPRAGE images often fail on SPGR images and underperform on MPRAGE images acquired by different scanners or with slightly different parameters~\cite{pham2020contrast}. 
Scanner software and calibration differences can also contribute to this effect.

The issue of contrast variation is commonly seen in multi-site studies, where a trained model degrades in performance when tested on data from another site (i.e., the \textit{domain shift} problem). 
This is because ML based methods assume the training and testing are conducted on data drawn from the same distribution~(domain). 
This is not the case for MR images acquired from different sites, scanners, or with differing imaging parameters. 
For example, T$_1$-w images acquired from two scanners with different configurations should obviously be treated as two domains. 
However, T$_1$-w and T$_2$-w images acquired from the same scanner should also be considered as coming from two domains. 

MR image harmonization~\cite{dewey2019deepharmony} alleviates domain shift by treating the problem as an \textit{image-to-image translation}~(IIT) (or synthesis) task, where the goal is to translate image contrasts between domains (e.g., T$_1$-w images acquired from different scanners). 
MR harmonization can be separated into two categories: supervised and unsupervised. 
In the supervised setting, MR images of the same anatomy across multiple sites are available; these are known as \textit{traveling subjects} or \textit{inter}-site paired data. 
These images are used to train intensity transformations between sites. 
However, traveling subjects are impractical in large-scale multi-site harmonization tasks. 
Unsupervised harmonization methods do not require inter-site paired data. 
Instead, these methods often rely on domain-specific models~(e.g., intensity transformations and discriminators).
We outline recent related work in IIT and unsupervised domain adaptation~(UDA), below.

IIT learns a transformation of images between different domains, e.g., MR to CT~\cite{zhao2017supervoxel,wolterink2017deep} or T$_1$-w to T$_2$-w~\cite{zuo2020synthesizing}. 
In both supervised and unsupervised IIT, the goal is to approximate the joint distribution drawn from the (two) domains.
Supervised IIT methods use pixel-to-pixel reconstruction error during model training. 
Recent unsupervised IIT work has explored learning disentangled representations~\cite{huang2018multimodal,lee2018diverse,xia2020unsupervised}, the idea being to tease apart the domain-invariant and domain-specific information in the representation. 
As an unsupervised IIT method, unsupervised harmonization faces four challenges. 
First, the lack of inter-site paired data along with the coupling theory~\cite{lindvall2002lectures} tells us there are infinitely many possible joint distributions given the two marginal distributions.
Therefore, to learn a meaningful harmonization model~(joint distribution), further constraints are required. 
Cycle-consistency is commonly assumed in unsupervised IIT~\cite{zhu2017unpaired,huang2018multimodal,liu2018unified}. 
However, there is no theory that supports the validity of this assumption.
Second, the lack of inter-site paired data means that pixel-to-pixel regularization cannot be easily achieved. 
Domain-specific discriminators are commonly used in many unsupervised IIT methods~\cite{huang2018multimodal,liu2018unified,zhu2017unpaired}. 
For harmonization, performance will heavily rely on the discriminators' ``judgement'' during training, and geometry shift is a common drawback in unsupervised harmonization. 
Third, the use of site specific-models means that the size of the harmonization model grows with the number of sites. 
Lastly, most existing harmonization approaches are not able to work on domains not seen in the training data. 
When testing and training domains differ, most methods require retraining with images from all domains, which is not practical.

In general, the goal of UDA is to learn a model from a source domain with labeled data and apply it to a target domain with unlabeled data during testing~\cite{saito2018maximum,he2020self,kamnitsas2017unsupervised,varsavsky2020test}. 
Without special design, domain shifts between training and testing can cause a performance drop.
Different from IIT, which aims at mapping image data across domains, a UDA tries to adjust the model during testing. 
A UDA allows the model to detect a domain and then provide self-supervision for model adjustment during testing. 
UDAs are especially helpful in medical imaging, where training and testing data are likely to come from different sources.

\begin{figure}[!tb]
    \centering
    \begin{tabular}{c c c}
        \includegraphics[width=0.37\textwidth]{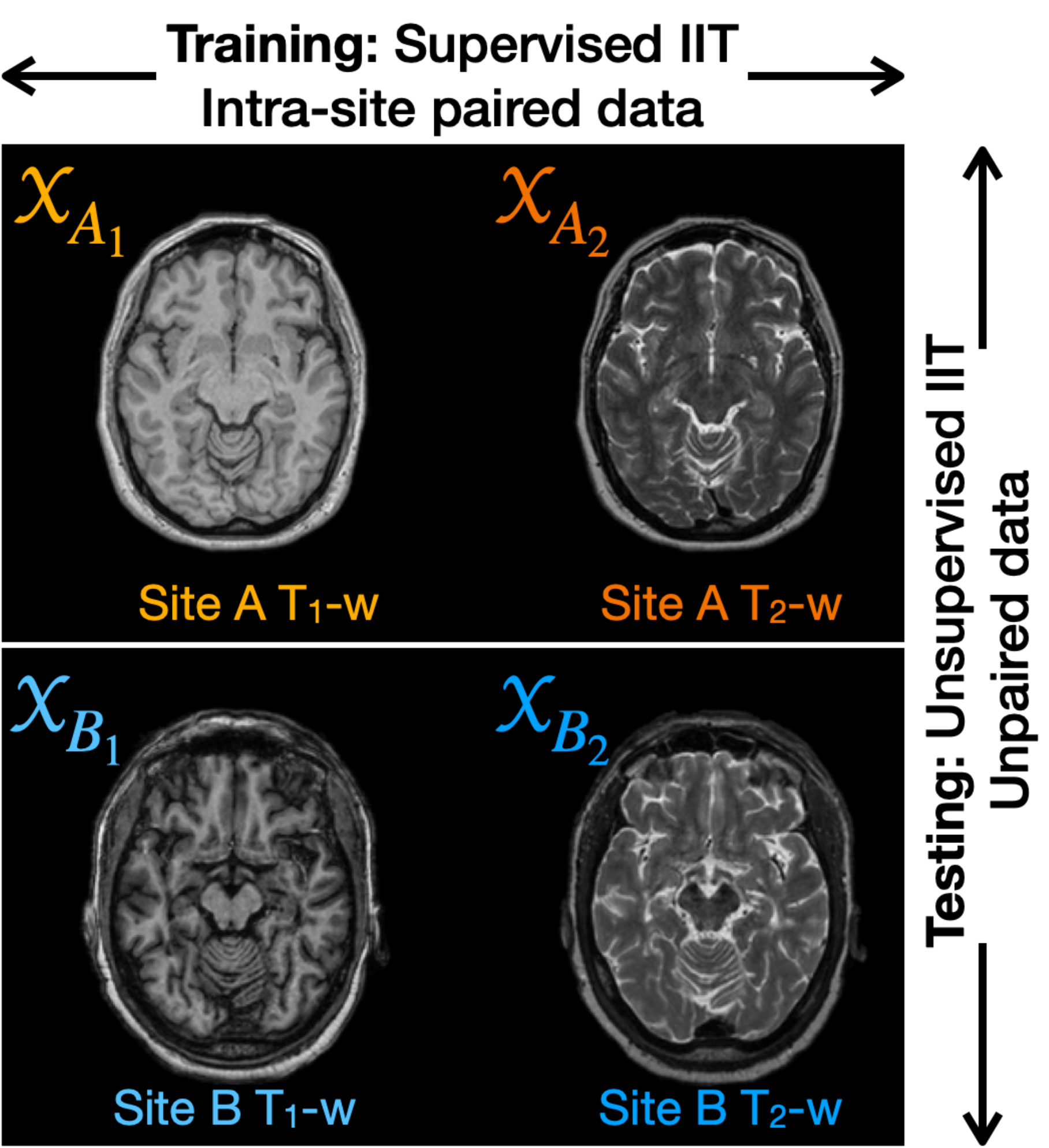}&&
        \includegraphics[width=0.60\textwidth]{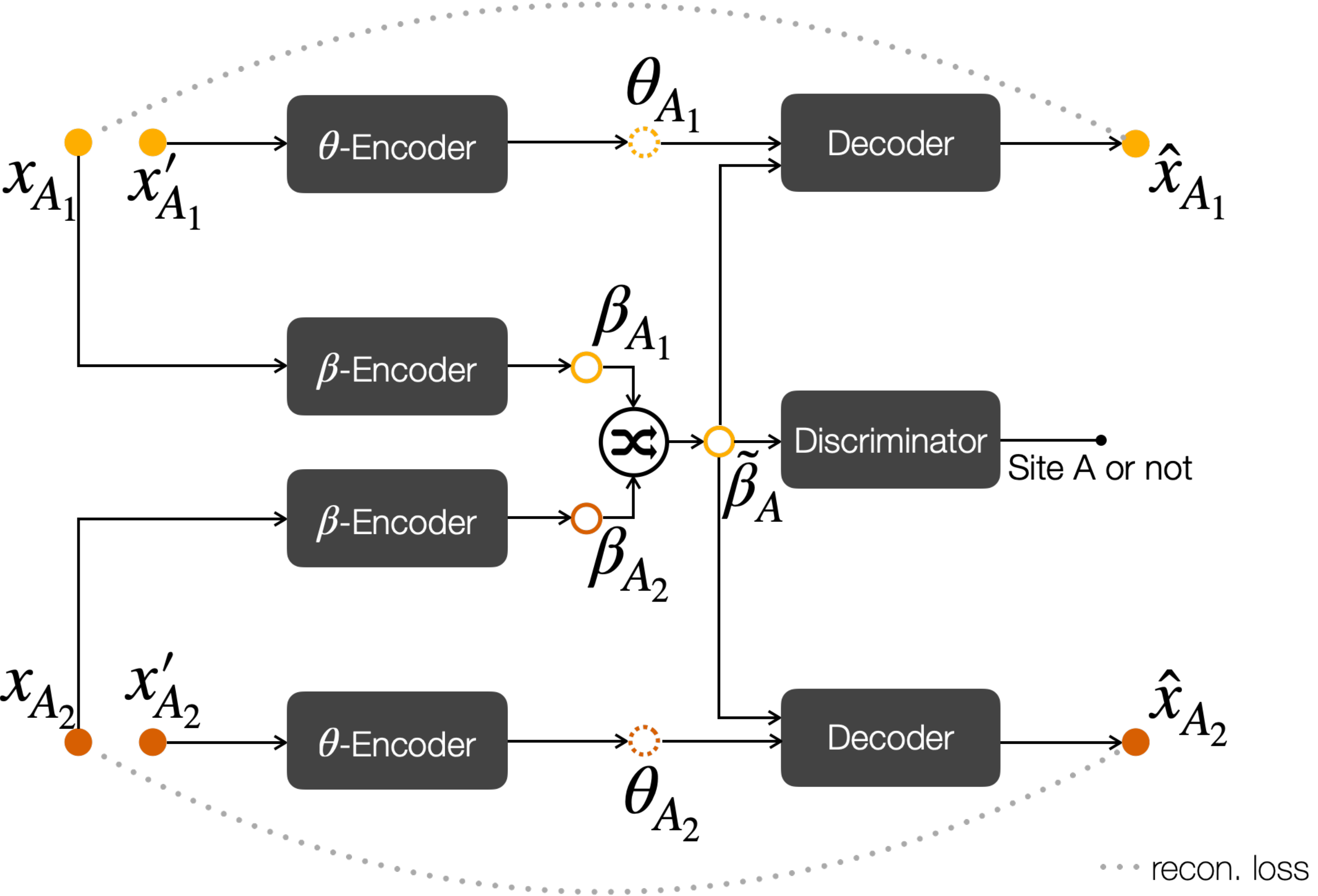} \\
        \textbf{(a)} & \hspace*{1em} &\textbf{(b) }
    \end{tabular}
    \caption{\textbf{(a)} Given T$_1$-w and T$_2$-w images from Sites $A$ and $B$,
our method solves intra-site supervised IIT~(T$_1$--T$_2$ synthesis) and
inter-site unsupervised IIT~(harmonization), where an alphabetical index indicates site and a numerical index indicate MR contrast. \textbf{(b)} The proposed method consists of a \textit{single} $\theta$-encoder, a $\beta$-encoder, a decoder, and a $\beta$-discriminator that work on all domains. $x$ and $x'$ share the same contrast but have different anatomy. The same networks work on all sites.}
    \label{fig:framework}
\end{figure}

We propose an unsupervised harmonization approach, CALAMITI (contrast anatomy learning and analysis for MR intensity translation and integration), which integrates the merits of both IIT and UDA. 
Building upon the recent work in~\cite{dewey2020disentangled}, we use the routinely acquired multi-contrast MR images \textit{within} each site~(called \textit{intra}-site paired data) during the same imaging session. 
However, as we discuss in Section~\ref{sec:beta_discriminator}, this technique alone does not provide a globally disentangled latent space and cannot be easily generalized to data from a new site. 
CALAMITI is an improved, theoretically grounded, unsupervised harmonization approach based on an information bottleneck~(IB)~\cite{tishby2000information} that learns a global, disentangled latent space of anatomical and contrast information and can be easily adapted to a new testing site using only the new data.
To our knowledge, this is the first work that overcomes the four challenges in unsupervised harmonization.
\textbf{First}, by taking advantage of the intra-site paired data, the proposed method solves an unsupervised IIT problem in a supervised way, avoiding introducing any extra constraint~(e.g., cycle-consistency) on the model and achieving better pixel-to-pixel regularization.
\textbf{Second}, it has a unified structure for multi-site harmonization, which means that model size does not grow with the number of sites.
\textbf{Third}, it provides a global latent space for all training data by encouraging a consistent description of the anatomy.
\textbf{Finally}, it is able to adapt to a new site without any retraining on the original data.
For all of this work, we also provide a theoretical explanation of the disentangled latent space using IB theory.

\section{Method}
\label{sec:method}
\subsection{The disentangling framework}
Our method uses multi-contrast MR images of the same subject \textit{within} each site~(intra-site paired data) to train a cross-site harmonization model. 
Here, we emphasize the relationship between ``site'', ``domain'', and ``MR contrast''.
As shown in Fig.~\ref{fig:framework}(a), given T$_1$-w and T$_2$-w images from Sites $A$ and $B$, there are four domains~$\mathcal{X}_{A_1}$, $\mathcal{X}_{A_2}$, $\mathcal{X}_{B_1}$, and $\mathcal{X}_{B_2}$, where an alphabetical index indicates site and a numerical index represents contrast. Our goal is to learn a disentangled representation that captures anatomical and contrast information from the input images. 
Following the notation in~\cite{dewey2020disentangled}, the anatomical representation~($\beta$) is domain-invariant and the contrast representation~($\theta$) has some domain-specific information. 
Thus, combining the $\beta$ from one site with the $\theta$ from another allows harmonization across sites. 
To learn the disentangled representation, we solve the inter-site unsupervised IIT problem based on training from intra-site supervised data.

Figure~\ref{fig:framework}(b) outlines our framework, which consists of a $\theta$-encoder, a $\beta$-encoder, a decoder, and a $\beta$-discriminator that work on all domains.
Here, we outline the high-level training strategy using the proposed framework, and we highlight the role of our $\beta$-discriminator in Section~\ref{sec:beta_discriminator}. 
Each site has paired T$_1$-w and T$_2$-w images---with different imaging parameters at each site---which train a disentangled network in a supervised IIT way. 
For example, intra-site paired images $x_{A_1}$ and $x_{A_2}$ of the same subject imaged at Site~$A$~(in our case T$_1$-w and T$_2$-w images from Site~$A$) are sent to a $\beta$-encoder to extract anatomical information. 
These images have the same anatomy, so the extracted anatomical representations $\beta_{A_1}$ and $\beta_{A_2}$ should be the same. 
To encourage similarity of $\beta$, we randomly shuffle between $\beta_{A_1}$ and $\beta_{A_2}$ before decoding as well as introduce a small $l_1$ loss between the two $\beta$'s. 
To prevent contrast representation $\theta$ from capturing anatomical information, we provide the $\theta$-encoder with an image of different anatomy (but the same contrast), $x_{A_1}'$. 
This is achieved by selecting a different slice from the same volume as $x_{A_1}$. 
The decoder takes the randomly selected anatomical representation~($\tilde{\beta}_A$), concatenated with a $\theta$ to generate a synthetic image. 
The contrast of the synthetic image depends on which $\theta$ has been chosen. 
The \textit{same} $\beta$-encoder, $\theta$-encoder, decoder, and $\beta$-discriminator are used for all training sites to achieve a unified structure.

Our $\beta$-encoder and decoder both have a U-Net like architecture with four downsampling layers, while the $\theta$-encoder is four convolutional layers followed by three fully connected layers.
$\beta$ is one-hot encoded with multiple channels and the same spatial extents as the input image. 
For gradients to backpropagate through the one-hot encoded $\beta$, we adopt and implement the trick introduced in~\cite{dewey2020disentangled, liu2020variational}, wherein $\beta$'s are calculated using a Gumbel-softmax layer. One-hot encoding $\beta$ restricts its capacity, encouraging $\beta$ to capture only anatomical information. 

\subsection{Creating a consistent anatomical space}
\label{sec:beta_discriminator}
To learn a consistent anatomical space for all sites, we introduce a \mbox{$\beta$-discriminator} to our framework. 
Because our training strategy only uses supervised IIT \textit{within} each site---with no supervision \textit{between} sites---the $\beta$-encoder could possibly learn a \textit{distinct} $\beta$ space for each site. 
In this case, the $\beta$'s and $\theta$'s are disentangled within each site, and we refer to it as a locally disentangled latent space. 
This is not desirable in harmonization, as combining these $\beta$'s and $\theta$'s across sites would not be ideal. 
To avoid this, we must encourage the learned $\beta$'s of all sites to be from the same distribution (i.e., $\beta$'s and $\theta$'s are globally disentangled).
This leads us to use a one-class discriminator on $\beta$ space to encourage distribution similarity. 
No matter which site an input $\beta$ comes from, the $\beta$-discriminator learns to distinguish whether the $\beta$ is from Site~$A$ or not, further pushing $\theta$ to describe the difference between sites as well as different MR contrasts. 

The proposed framework solves a number of outstanding problems. 
First, it performs unsupervised IIT by using supervised IIT during training. 
This avoids geometry issues inherent in unsupervised IIT by penalizing pixel-to-pixel error during training our framework. 
This is more effective than introducing a cycle consistency constraint, as cycle consistency still allows a model to learn ``circle--square--circle''.
Second, our unified harmonization structure means we have one $\beta$-encoder, one $\theta$-encoder, one decoder, and one $\beta$-discriminator that work on all domains; Section~\ref{sec:expt} includes an $8$-site experiment using this structure. 
This saves a significant number of parameters when there are many sites---e.g., the CycleGAN~\cite{zhu2017unpaired} requires $O(N^2)$ image translation models and $N$ discriminators, for $N$ sites. 
Third, our discriminator functions on the latent variables $\beta$ instead of the harmonized images. 
This encourages our decoder to act like a \textit{universal} imaging equation, generating synthetic MR images based on any $\beta$ and $\theta$ within the distribution of the training data. 
This strategy combined with the unified structure, makes our model more robust when there are more sites involved during training. 
Lastly, our discriminator makes a one-class decision: whether an input $\beta$ is from Site~$A$ or not. 
We show in Section~\ref{sec:da}, that this allows our model to adapt to a new testing site after fine tuning. 
Table~\ref{tab:comparison} provides a summary comparison of the proposed method with other unsupervised IIT approaches.

\begin{table}[!tb]
    \centering
    \caption{Features of recent unsupervised IIT and UDA approaches.}
    \resizebox{0.85\columnwidth}{!}{
    \begin{tabularx}{\textwidth}{C  C  C  C  C C  C}
    \toprule
        & \textbf{{Bidirection}} & \textbf{{Multiple domains}} & \textbf{Unified structure} & \textbf{Disentangle} & \textbf{Global latent space} & \textbf{Domain adaptation} \\
    \hline 
        {CycleGAN}~\cite{zhu2017unpaired}  & \ding{51}  & -- & -- & -- & -- & --  \\
        {UNIT}~\cite{liu2017unsupervised}  & \ding{51}  & -- & -- & -- & -- & --  \\
        {MUNIT}~\cite{huang2018multimodal} & \ding{51}  & \ding{51}  & -- & \ding{51} & -- & --\\
        {DCMIT}~\cite{xia2020unsupervised} & \ding{51}  & \ding{51}  & -- & \ding{51} & \ding{51} & -- \\
        {SDAnet}~\cite{he2020self}  & -- & -- & \ding{51}  & -- & -- & \ding{51}  \\
        {Dewey~et~al.}~\cite{dewey2020disentangled} & \ding{51}  & \ding{51}  & \ding{51}  & \ding{51}  & -- & -- \\
        {CALAMITI}   & \ding{51} & \ding{51} & \ding{51} & \ding{51} & \ding{51} & \ding{51} \\
    \bottomrule
    \end{tabularx}}
    \label{tab:comparison}
\end{table}

\subsection{Learning from an information bottleneck}
\label{sec:infomation}

By providing the $\theta$-encoder with an image of a different anatomy (but the same contrast) as provided to the $\beta$-encoder, we create a conditional variational autoencoder~(CVAE)~\cite{sohn2015learning} even though the condition variable $\beta$ is not connected to the $\theta$-encoder. 
All of these strategies help us to limit the information that can be passed through each of the $\beta$ and $\theta$ channels, which we now show theoretically forms an IB given the model design.

IB theory~\cite{tishby2000information} describes a constrained optimization problem with the goal of learning a compressed latent representation $Z$ such that the mutual information~(MI) between $Z$ and the task variable $Y$ is maximized while $Z$ captures minimum information about the input variable $X$. 
Mathematically, this can be formulated as $Z^* = \argmin_Z I(Z;X) - \lambda I(Z;Y)$, where $I(\cdot;\cdot)$ is the MI and $\lambda$ is a hyper-parameter. 
IB theory is closely related to the variational autoencoders~(VAEs)
and disentangled representation learning~(cf.~\cite{burgess2018understanding,dai2018compressing}).
\cite{alemi2016deep} showed that IB is a more general case of the VAE
objective.

Our network structure forms a CVAE. 
To better illustrate the IB in a general setting, we slightly modify the notation~(i.e., remove site index), and highlight the CVAE structure.
\begin{theorem}
It can be shown that optimizing our network structure is equivalent to
solving a conditional IB problem,~i.e.,
\begin{equation}
    \theta^* = \argmin_\theta I(X';\theta) - \lambda
I(X;\theta|\tilde{\beta}).
\label{eq:information}
\end{equation}
\label{theorem:information}
\end{theorem}
The proof of Theorem~\ref{theorem:information} is similar to that in~\cite{alemi2016deep}, despite the fact that we are solving a CVAE problem with condition $\tilde{\beta}$. 
The reason why the first term of Equation~\ref{eq:information} is free from condition $\tilde{\beta}$ is because $\tilde{\beta}$ is disentangled from $\theta$ and will be ignored in calculating $\theta$. 
An intuitive understanding of Equation~\ref{eq:information} is that the proposed method learns a contrast representation $\theta$ that captures minimum information about the input variable $X'$, while the (conditional) MI between $\theta$ and the target variable $X$ is maximized. 
Since the shared information between variables $X$ and $X'$ is the contrast, we would expect $\theta$ to capture only contrast information about $X'$ after training. Equation~\ref{eq:information} can be re-organized as a KL divergence term and a reconstruction term~(similar to the CVAE loss~\cite{sohn2015learning}), and directly optimized as network loss functions, i.e., 
\begin{equation}
    \theta^* = \argmin_\theta \mathcal{D}_{\text{KL}}\left[
p(\theta|x') || p(\theta) \right] - \lambda
\mathbb{E}_{p(\theta|x')} \left[ \log
p(x|\theta,\tilde{\beta}) \right],
\end{equation}
where $p(\theta)$ is a zero mean unit variance Gaussian distribution. 
$p(\theta|x')$ and $p(x|\theta,\tilde{\beta})$ can be modeled by a probabilistic $\theta$-encoder and decoder, respectively. 
This KL divergence term encourages a bounded value for $\theta$, which even if lightly weighted restricts its possible expression. 
Accordingly, our network loss functions include a reconstruction loss for supervised IIT~($l_1$ and perceptual loss~\cite{johnson2016perceptual}), a KL divergence term on $\theta$, an adversarial loss between $\beta$-discriminator and $\beta$-encoder, and a similarity loss on $\beta$ between contrasts\footnote{Code is available at \url{https://iacl.ece.jhu.edu/index.php?title=CALAMITI}}. 

\subsection{Domain adaptation}
\label{sec:da}

Suppose the proposed model was pretrained on Sites~$A$ and $B$, and the goal is to harmonize a new site, Site~$C$, to Site~$A$ or $B$ without a retraining that includes data from all sites. As in regular CALAMITI training, the supervised IIT is conducted on images from Site~$C$. 
However, the decoder and $\beta$-discriminator weights are frozen, and only the last few layers of the $\beta$- and $\theta$-encoders are updated. 
We rely on the assumption that our decoder is well-generalized in previous training to produce a variety of contrast images. 
Our $\beta$-discriminator guides the $\beta$-encoder to generate $\beta$'s that follow the previously learned distribution of $\beta$; avoiding a Site~$C$ specific $\beta$ space. 
Thus our $\beta$-discriminator acts as a domain shift detector on $\beta$ space like other UDA methods~\cite{he2020self,varsavsky2020test}. 
Until the $\beta$-encoder generates $\beta$ for Site~$C$ that is less distinguishable from previously learned $\beta$, the $\beta$-discriminator will produce a loss for mismatching $\beta$'s. 
As the decoder weights are frozen in fine tuning, once $\beta$'s from Site~$C$ match previously learned $\beta$ distribution, we are able to harmonize images between new and previous
sites.  
Thus by combining $\beta_{C_1}$ with $\theta_{A_1}$, we can harmonize anatomy imaged at Site~$C$ with the corresponding contrast from Site~$A$. 
Thus Site~$C$ can be harmonized to any site included in the original training.

\section{Experiments and Results}
\label{sec:expt}
\subsection{Datasets and preprocessing}
\begin{table}[!tb]
    \centering
    \caption{Scanner make, image sequence and parameters~(TE, TR, TI if necessary), and
acquisition extent. T$_1$-w sequence key: M - MPRAGE; ME - MEMPRAGE.}
    \resizebox{0.99\columnwidth}{!}{
    \begin{tabular}{lcccc}
        \cmidrule[0.35mm]{1-5}
        \multirow{2}{*}{ } 
            & \textbf{Site~$\boldsymbol{A}$}~(IXI~\cite{ixi}) 
            & \textbf{Site~$\boldsymbol{B}$}~(IXI) 
            & \textbf{Site~$\boldsymbol{C}$}~(OASIS3~\cite{LaMontagne}) 
            & \textbf{Site~$\boldsymbol{D}$}~(OASIS3) \\
            & Philips Intera 1.5T 
            & Philips Gyroscan 3.0T 
            & Siemens Sonata 1.5T
            & Siemens TimTrio 3.0T \\
            \cmidrule(lr){1-5}
        \multirow{2}{*}{\textbf{T$_1$-w}}
            & M: 4.6ms, unknown, unknown 
            & M: 4.6ms, unknown, unknown
            & M: 3.93ms, 1.9s, 1.1s 
            & M: 3.16ms, 2.4s, 1s \\
            & $1.2\times0.94\times0.94$mm 
            & $1.2\times0.94\times0.94$mm
            & $1\times1\times1$mm 
            & $1\times1\times1$mm \\
            \cmidrule(lr){1-5}
        \multirow{2}{*}{\textbf{T$_2$-w}}
            & TSE: 100ms, 8.2s 
            & TSE: 100ms, 8.2s
            & TSE: 116ms, 6s 
            & TSE: 455ms, 3.2s\\
            & $0.94\times0.94\times1.25$mm 
            & $0.94\times0.94\times1.25$mm
            & $0.9\times0.9\times5$mm 
            & $1\times1\times1$mm \\
            \cmidrule[0.35mm]{1-5}
        \multirow{2}{*}{ } 
            & \textbf{Site~$\boldsymbol{E}$}~(OASIS3) 
            & \textbf{Site~$\boldsymbol{F}$}~(OASIS3) 
            & \textbf{Site~$\boldsymbol{G}$}~(Private) 
            & \textbf{Site~$\boldsymbol{H}$}~(Private) \\
            & Siemens TimTrio 3.0T 
            & Siemens BioGraph 3.0T
            & Philips Achieva 3.0T 
            & Philips Achieva 3.0T \\
            \cmidrule(lr){1-5}
        \multirow{2}{*}{\textbf{T$_1$-w}}
            & M: 3.16ms, 2.4s, 1s 
            & M: 2.95ms, 2.3s, 0.9s
            & ME: 6.2ms, 2.5s, 0.9s 
            & M: 6ms, 3s, 0.84s \\
            & $1\times1\times1$mm 
            & $1.05\times1.05\times1.2$mm
            & $1\times1\times1$mm 
            & $1.1\times1.1\times1.18$mm \\
            \cmidrule(lr){1-5}
        \multirow{2}{*}{\textbf{T$_2$-w}}
            & TSE: 455ms, 3.2s
            & TSE: 454ms, 3.2s 
            & TSE: 240ms, 2.5s 
            & TSE: 80ms, 4.2s \\
            & $1\times1\times1$mm
            & $1\times1\times1$mm
            & $1\times1\times1$mm 
            & $1.1\times1.1\times2.2$mm \\
            \cmidrule[0.35mm]{1-5}
        \end{tabular}}
    \label{tab:scanners}
\end{table}

MR images, both T$_1$-w and T$_2$-w, acquired from eight different sites~(scanners) were used to evaluate the proposed method. 
Data source, scanner details, and imaging parameters are provided in Table~\ref{tab:scanners}. 
Sites~$A$ thru~$F$ are healthy controls, Sites~$G$ and~$H$ imaged multiple sclerosis~(MS) subjects. 
Sites~$D$ and~$E$ have similar scanners and sequences and thus similar contrast. 
Images underwent preprocessing including N4 inhomogeneity correction, super-resolution for 2D acquired scans, registration to $1~\text{mm}^3$ MNI space, and white matter~(WM) peak normalization. 
The center $60$ axial slices with spatial dimension of $224\times192$ were extracted for each subject.

\subsection{Qualitative and quantitative evaluation}
\begin{figure}[!tb]
    \centering
    \includegraphics[width=\textwidth]{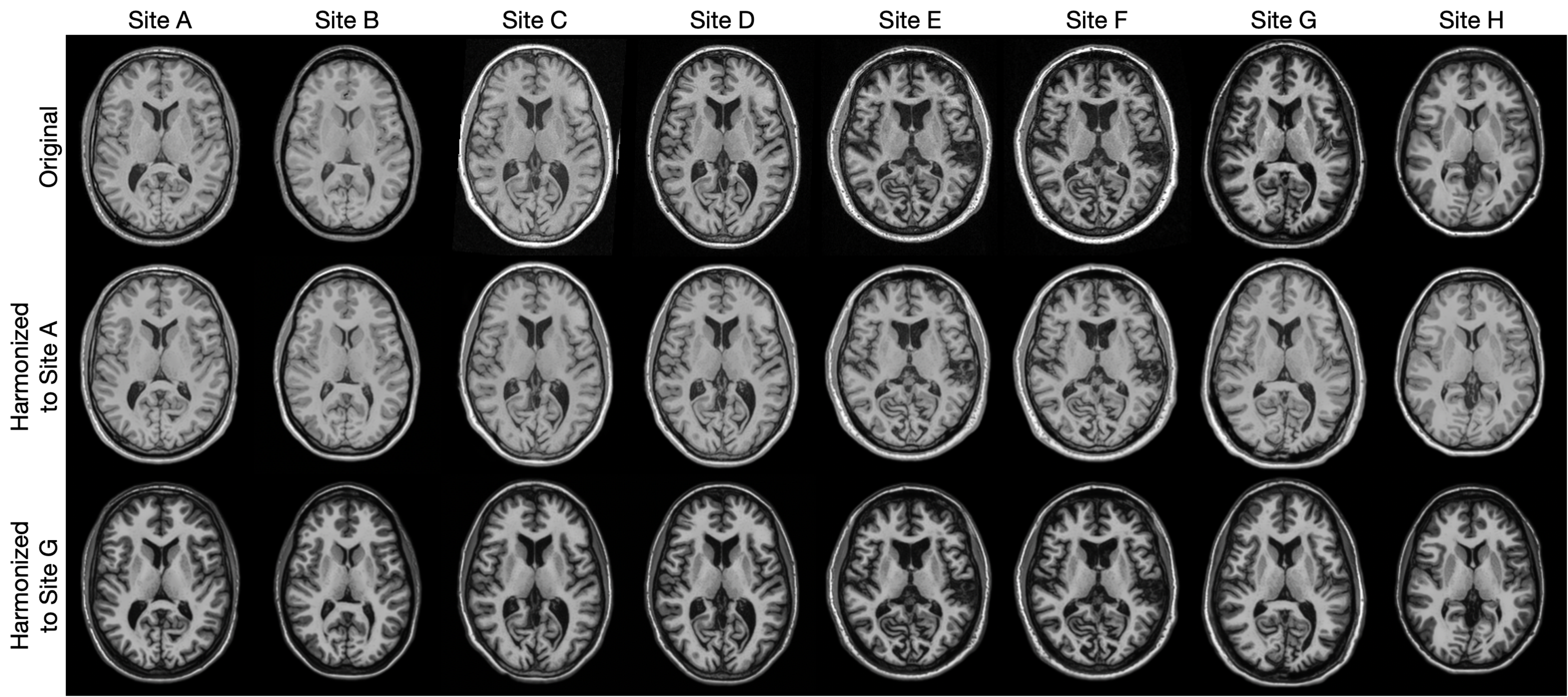}
    \caption{Harmonization results of the proposed method. T$_1$-w MR images from eight sites with different manufacturer and imaging parameters are harmonized to Site~$A$~(middle row) and Site~$G$~(bottom row). The contrast of harmonized images is determined by the mean $\theta$ value over all testing images at a site. }
    \label{fig:harmonized_imgs}
\end{figure}

\begin{figure}[!tb]
    \centering
    \includegraphics[width=0.7\textwidth]{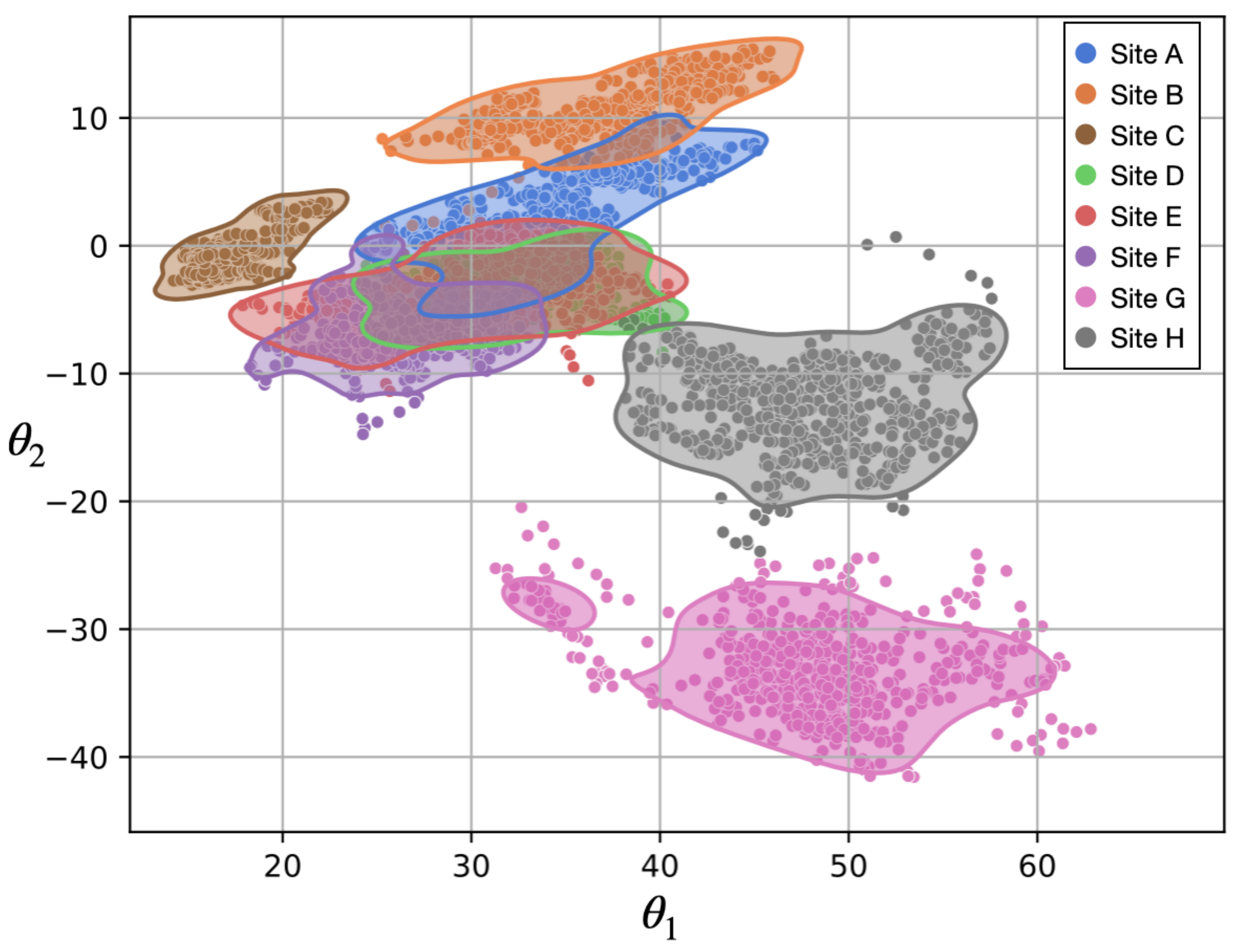}
    \caption{Visualization of $\theta$ space on testing T$_1$-w images. Contours are fitted to the $\theta$ clusters for visualization purpose.}
    \label{fig:theta_space}
\end{figure}

For Sites~$C$, $D$, $E$, and~$F$, there are 10 subjects~(600 axial slices) used from each site for training.
For the remaining sites, 20 subjects from each site are used in training. 
There are longitudinal scans in the OASIS3~\cite{LaMontagne} dataset acquired by different scanners with a short period between visits. 
These scans are held-out and used as traveling subjects for quantitative evaluation in testing.
Specifically, there are seven traveling subjects between Site~$C$ and $D$, and ten traveling subjects between Site~$E$ and Site~$F$.
The average days between two visits for Sites~$C/D$ and Sites~$E/F$ are 162~days and 13~days, respectively. 
In our experiments, $\beta$ is a four-channel one-hot encoded map, with spatial dimension the same as the image, while $\theta$ is a two-dimensional vector.
Figure~\ref{fig:harmonized_imgs} shows harmonized MR images from the eight sites. 
The mean $\theta$ value of all testing images at each site was used to harmonize images between sites. 
Our analysis has focused on the T$_1$-w images, as these represent the images with the greatest disparity across the imaging sites and the primary contrast for neuroimaging analysis. 
With regard to the T$_2$-w images, we achieve similar image quality as the T$_1$-w images. Figure~\ref{fig:theta_space} shows $\theta$ values of the held-out T$_1$-w testing images.
We observe that Sites~$D$ and $E$ overlap, which is good as the sites have identical scanner configurations.
Second, images acquired using different methods~(MPRAGE, and MEMPRAGE) are separated.
Third, the distance in $\theta$ space corresponds to human perception; sites with visually similar contrast have closer clusters in $\theta$ space. 
For example, although images in Sites~$D$, $E$, $F$, and $H$ are all MPRAGE images, $\theta$ points are more closely clustered in Sites~$D$, $E$ and $F$, than Site~$H$. 

\begin{figure}[!tb]
    \centering
    \includegraphics[width=0.85\textwidth]{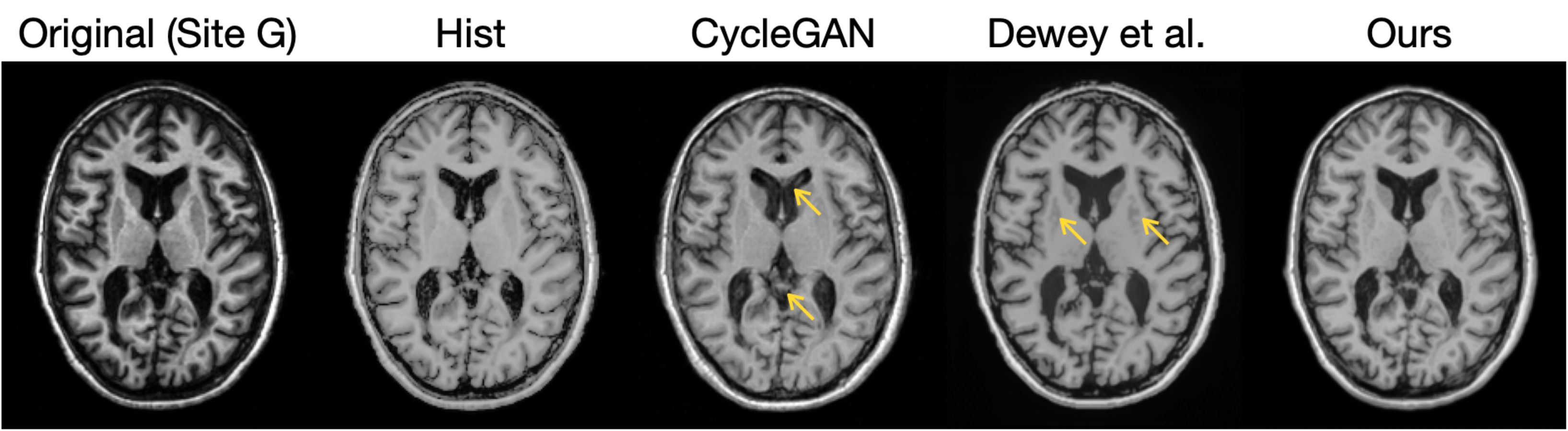}
    \caption{Visual comparison of different harmonization approaches. An MR image from Site~$G$ is harmonized to Site~$A$~(see Fig.~\ref{fig:harmonized_imgs} for reference images). Yellow arrows indicate geometry change.}
    \label{fig:compare_methods}
\end{figure}

In Table~\ref{tab:quantitative} and Fig.~\ref{fig:compare_methods}, we show qualitative and quantitative comparison of different unsupervised IIT methods. 
The traveling subjects are used in the quantitative comparison. 
Specifically, our baseline is the MR images without harmonization~(No har), and we compare the structural similarity index measurement~(SSIM) and peak signal-to-noise ratio~(PSNR) of histogram matching~(Hist), CycleGAN~\cite{zhu2017unpaired}, Dewey~et~al.~\cite{dewey2020disentangled}, and CALAMITI. 
Histogram matching is a non-training method, while the other approaches are ML-based. 
To select a reference image for histogram matching, we first randomly chose a volume, then selected the same slice number as our source image. 
For a fair comparison, we consider two training scenarios for CALAMITI: only include two sites~(the source and target site in harmonization) or include all eight sites during training. 
Paired Wilcoxon signed rank tests were conducted between CALAMITI~(two sites) and each comparison method under each performance measurement. 
Results show that CALAMITI has significantly~($p<0.001$, $N=420$ for Sites~$C$ and $D$, $N=600$ for Sites~$E$ and $F$) better performance over all comparison methods, except for the PSNR of Site~$D\rightarrow C$. The null hypothesis is that the difference of SSIM or PSNR between the two sites is from a distribution with zero median.
Interestingly, CALAMITI has slightly better performance when more sites are used in training.
We suggest two possible reasons for this. 
First, CALAMITI has a unified structure, which makes the whole model less likely to overfit. 
Second, our $\beta$-discriminator is a single class discriminator, so more sites should improve the robustness of the discriminator.

\begin{table}[!tb]
    \centering
    \caption{Numerical comparison (Mean$\pm$Std. Dev.) of unsupervised IIT approaches. From left to right: no harmonization~(No har), histogram matching~(Hist), CycleGAN~\cite{zhu2017unpaired}, Dewey~et~al.~\cite{dewey2020disentangled}, and the proposed method~(Ours). The proposed method shows significant improvements over all comparison methods based on paired Wilcoxon signed rank tests, with an exception of PSNR of Site~$D \rightarrow C$. Bold numbers indicate the best mean performance.}
    \resizebox{\columnwidth}{!}{
    \begin{tabular}{c|c|cccccc}
        \toprule 
    & & \textbf{No har} & \textbf{Hist} & \textbf{CycleGAN} & \textbf{Dewey~et~al.} & \textbf{Ours~(2 sites)} & \textbf{Ours~(8 sites)}\\
        \hline 
        \multirow{2}{*}{\textbf{Site~C$\rightarrow$D}} 
        & {SSIM} & $0.8034\pm 0.0184$ & $0.8349\pm 0.0456$ & $0.8729\pm 0.0346$ & $0.8637\pm 0.0345$ & $0.8811\pm0.0292$ & $\boldsymbol{0.8814\pm 0.0254}$ \\
        & {PSNR} & $26.81\pm 1.07$ & $28.03\pm 1.45$ & $29.60\pm 1.62$ & $29.35\pm 1.06$ & $29.80\pm0.98$ & $\boldsymbol{29.82\pm 0.80}$ \\
        \hline
        \multirow{2}{*}{\textbf{Site~D$\rightarrow$C}} 
        & {SSIM} & $0.8034\pm 0.0184$ & $0.7983\pm 0.0297$ & $0.8583\pm 0.0264$ & $0.8554\pm 0.0300$ & $0.8617\pm0.0245$&  $\boldsymbol{0.8663\pm 0.0233}$ \\
        & {PSNR} & $26.81\pm 1.07$ & $27.53\pm 1.21$ & ${28.63\pm 1.52}$ & $28.31\pm 1.30$ & ${28.50\pm1.20}$ & $\boldsymbol{28.68\pm 1.28}$ \\
        \hline
        \multirow{2}{*}{\textbf{Site~E$\rightarrow$F}} 
        & {SSIM} & $0.8706\pm 0.0447$ & $0.8255\pm 0.0375$ & $0.8737\pm 0.0404$ & $0.8748\pm 0.0400$ & $0.8815\pm0.0366$ &$\boldsymbol{0.8834\pm 0.0374}$ \\
        & {PSNR} & $29.74\pm 1.72$ & $27.46\pm 1.04$ & $29.70\pm 2.16$ & $29.66\pm 1.82$ & $30.05\pm1.72$ & $\boldsymbol{30.17\pm 1.82}$ \\
        \hline
        \multirow{2}{*}{\textbf{Site~F$\rightarrow$E}} 
        & {SSIM} & $0.8706\pm 0.0447$ & $0.7935\pm 0.0596$ & $0.8719\pm 0.0537$ & $0.8719\pm 0.0482$ & $0.8817\pm0.0402$ &$\boldsymbol{0.8843\pm 0.0464}$ \\
        & {PSNR} & $29.74\pm 1.72$ & $27.75\pm 1.26$ & $29.79\pm 1.92$ & $29.54\pm 1.57$ & $30.28\pm1.55$ & $\boldsymbol{30.36\pm 1.68}$ \\
        \bottomrule
    \end{tabular}}
    \label{tab:quantitative}
\end{table}

\subsection{Domain adaptation}
\label{sec:da_results}
\begin{figure}[!tb]
    \centering
    \includegraphics[width=0.8\textwidth]{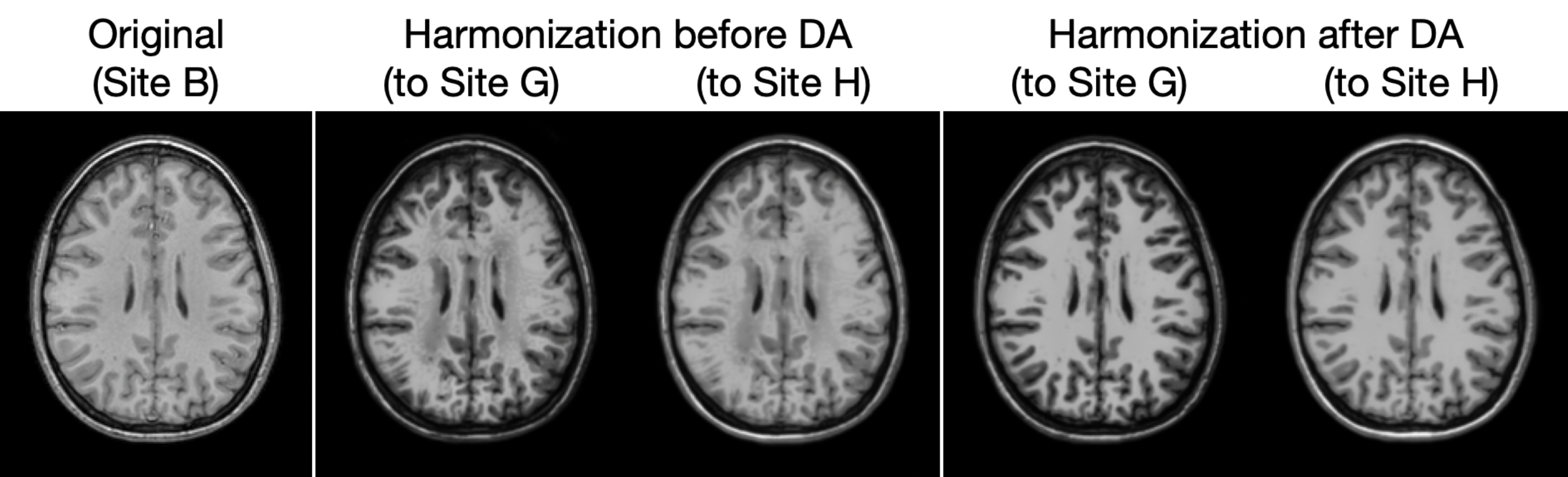}
    \caption{Visual comparison of fine tuning. Harmonization model is trained on Sites~$G$ and $H$, and tested on Site~$B$. The task is to harmonize Site~$B$ images to Site~$G$ or Site~$H$. Fine tuning is conducted only on Site~$B$.}
    \label{fig:fine_tune_viz}
\end{figure}

We provide both qualitative and quantitative results to test domain adaptation~(DA). 
For qualitative comparison, we trained our harmonization model using Sites~$G$ and $H$ and applied the trained model on Site~$B$ in testing. 
The task is to harmonize Site~$B$ to Site~$G$ or Site~$H$ without using data from Sites~$G$ or $H$. 
As shown in Fig.~\ref{fig:fine_tune_viz}, after fine tuning only on a subset of Site~$B$ images, the model is able to adjust itself to produce reasonable harmonization results. 
Table~\ref{tab:da} shows quantitative DA results. 
In each experiment, the proposed method was trained on the corresponding target site and Site~$G$, while testing and fine tuning were conducted solely on the source site. 
For example, when evaluating DA in Site~$C\rightarrow D$, the model was trained on Site~$D$ and $G$, and Site~$C$ was used as a previously unseen site for testing and fine tuning. 
Results show that the proposed method achieves significant~($p<0.001$, $N=420$) improvements after DA. 

\begin{table}[!tb]
    \centering
    \caption{Demonstration of domain adaptation feature of our method. Paired Wilcoxon signed rank tests show that the proposed method achieves significant~($p<0.001$, $N=420$) improvements after DA. }
    \resizebox{0.8\columnwidth}{!}{
    \begin{tabular}{c|cccc|ccc}
    \toprule
         & & \multicolumn{3}{c|}{\textbf{Site~C$\rightarrow$D}} &
\multicolumn{3}{c}{\textbf{Site~D$\rightarrow$C}} \\
         & & \textbf{Before DA} &\hspace*{2em}& \textbf{After DA} & \textbf{Before
DA} &\hspace*{2em}& \textbf{After DA} \\
    \hline
        \textbf{SSIM} && $0.8729\pm0.0301$ && $0.8743\pm0.0291$
              & $0.8028\pm0.0309$ && $0.8486\pm0.0253$ \\
        \textbf{PSNR} && $29.01\pm1.07$ && $29.41\pm0.98$
              & $25.37\pm1.31$ && $27.82\pm0.82$ \\
    \bottomrule
    \end{tabular}}
    \label{tab:da}
\end{table}

\section{Discussion and Conclusion}
Both qualitative and quantitative results from our eight-site experiment show the potential of the proposed method in large-scale multi-site studies. There are some limitations. First, the requirement of intra-site paired images in training could potentially restrict some applications---pediatric data for example---where acquiring multi-contrast images is not practical. Second, in our experiments, we only used paired T$_1$-w and T$_2$-w images. However, the proposed method can be extended to include more contrast MR images such as fluid-attenuated inversion recovery~(FLAIR) images to achieve a better disentanglement. Third, our experiments on MS patients show that the proposed method does not produce a satisfactory harmonization result on WM lesion areas. We hypothesize that inclusion of FLAIR images would improve this. Fourth, although satisfactory results have been observed in Section~\ref{sec:da_results}, the way we used our $\beta$-discriminator to update the $\beta$-encoder during domain adaptation is theoretically flawed. According to Goodfellow~et~al.~\cite{goodfellow2020generative}, the generator and discriminator must be updated jointly to achieve optimal performance. We view all these limitations as opportunities for future improvements.

In conclusion, we propose an unsupervised MR harmonization approach, CALAMITI, which integrates merits from both unsupervised IIT and UDA, and is grounded in information bottleneck theory. Our model learns a \textit{universal} imaging equation and a disentangled latent space without inter-site paired data. In contrast to many unsupervised harmonization methods, our model takes advantages of the intra-site paired data to prevent the geometry shift problem. Experiments show that the proposed approach achieves state-of-the-art harmonization performance both visually and in terms of SSIM and PSNR.

\section{Acknowledgments}
This research was supported by the TREAT-MS study funded by the Patient-Centered Outcomes Research Institute~PCORI/MS-1610-37115, the Intramural Research Program of the NIH, National Institute on Aging, and NIH grant R01-NS082347. 

\bibliographystyle{splncs04}
\bibliography{ml}

\end{document}